\documentclass[]{pasj02} 

\jyear{2025}
\Received{2024/12/10}%{yyyy/mm/dd}
\Accepted{2025/03/18}%{yyyy/mm/dd}
\usepackage[switch,mathlines]{lineno} % add line number to manuscript
\usepackage{natbib}

\def\OIII{[\mbox{O\,{\sc iii}}]$\lambda 5007$}

\def\OIIIab{[\mbox{O\,{\sc iii}}]$\lambda\lambda 4959,5007$}
\def\SIIab{[\mbox{S\,{\sc ii}}]$\lambda\lambda 6717,6731$}

\def\NII{[\mbox{N\,{\sc ii}}]$\lambda 6584$}

\def\NIIab{[\mbox{N\,{\sc  ii}}]$\lambda \lambda 6547,6584$}

\def\OIIIb{[\mbox{O{\sc iii}}]$\lambda 5007$}

\def\Ha{{H$\alpha$}}
\def\Hb{{H$\beta$}}

\def\NIIHa{[\mbox{N\,{\sc ii}}]$\lambda 6583$/H$\alpha$}

\def\OIIIHb{[\mbox{O\,{\sc iii}}]$\lambda 5007$/H$\beta$}

\def\ergs{${\rm erg}~{\rm s}^{-1}$}
\def\kms{${\rm km}~{\rm s}^{-1}$}
\definecolor{myred}{rgb}{0.8, 0, 0}

\title{ AGN outflows and their properties in Mrk 766 as revealed by KOOLS-IFU on the Seimei Telescope}

\author{
 Kyuseok \textsc{Oh}				\altaffilmark{1}\altemailmark \email{oh@kasi.re.kr}, 
 Yoshihiro \textsc{Ueda}			\altaffilmark{2},
 Satoshi 	\textsc{Yamada}		\altaffilmark{3},
 Yoshiki 	\textsc{Toba}			\altaffilmark{4,5,6$^{\dag}$},
 Keisuke 	\textsc{Isogai}			\altaffilmark{7,8},  
 Atsushi 	\textsc{Tanimoto}		\altaffilmark{9},
 Shoji	\textsc{Ogawa}			\altaffilmark{10},
 Ryosuke	\textsc{Uematsu}		\altaffilmark{2},
 Yuya	\textsc{Nakatani}		\altaffilmark{2},
 Kanta	\textsc{Fujiwara}		\altaffilmark{2},
 Yuta	 	\textsc{Okada}			\altaffilmark{2},
 Kazuya 	\textsc{Matsubayashi}	\altaffilmark{11},
 Kenta	\textsc{Setoguchi}		\altaffilmark{2}
}
\altaffiltext{1}{Korea Astronomy and Space Science Institute, Daedeokdae-ro 776, Yuseong-gu, Daejeon 34055, Republic of Korea}
\altaffiltext{2}{Department of Astronomy, Kyoto University, Kitashirakawa-Oiwake-cho, Sakyo-ku, Kyoto 606-8502, Japan}
\altaffiltext{3}{RIKEN Cluster for Pioneering Research, 2-1 Hirosawa, Wako, Saitama 351-0198, Japan}
\altaffiltext{4}{National Astronomical Observatory of Japan, 2-21-1 Osawa, Mitaka, Tokyo 181-8588, Japan}
\altaffiltext{5}{Academia Sinica Institute of Astronomy and Astrophysics, 11F of Astronomy-Mathematics Building, AS/NTU, No.1, Section 4, Roosevelt Road, Taipei 10617, Taiwan}
\altaffiltext{6}{Research Center for Space and Cosmic Evolution, Ehime University, 2-5 Bunkyo-cho, Matsuyama, Ehime 790-8577, Japan}
\altaffiltext{7}{Okayama Observatory, Kyoto University, 3037-5 Honjo, Kamogatacho, Asakuchi, Okayama 719-0232, Japan}
\altaffiltext{8}{Department of Multi-Disciplinary Sciences, Graduate School of Arts and Sciences, The University of Tokyo, 3-8-1 Komaba, Meguro, Tokyo 153-8902, Japan}
\altaffiltext{9}{Graduate School of Science and Engineering, Kagoshima University, Kagoshima 890-0065, Japan}
\altaffiltext{10}{Institute of Space and Astronautical Science (ISAS), Japan Aerospace Exploration Agency (JAXA), 3-1-1 Yoshinodai, Chuo-ku, Sagamihara, Kanagawa 252-5210, Japan}
\altaffiltext{11}{Institute of Astronomy, School of Science, The University of Tokyo 2-21-1 Osawa, Mitaka, Tokyo 181-0015, Japan}

\footnotetext[$^\dag$]{NAOJ fellow}


\maketitle

\begin{document} 
% ----------------------------------------------------- 
% 		ABSTRACT
% ----------------------------------------------------- 
\begin{abstract}
We present the emission-line flux distributions and their ratios, as well as the gas outflow features, 
of the innermost 2 kpc region of the type 1 Seyfert galaxy Mrk 766, 
using the Kyoto Okayama Optical Low-dispersion Spectrograph 
with an optical-fiber integral field unit on the Seimei Telescope. 
We find that the central region of Mrk 766 is kinematically disturbed, 
exhibiting asymmetric and radially distributed AGN-driven ionized gas outflows traced by \OIII\ with velocities exceeding 500 \kms. 
The mass of the ionized gas outflow is estimated to be $10^{4.65-5.95} M_{\odot}$, 
and the mass outflow rate is $0.14-2.73$ M${\odot}$ yr$^{-1}$. 
This corresponds to a kinetic power, 
$\dot{E}_{\rm K}$, of $4.31 \times 10^{40} \ {\rm erg} \ {\rm s^{-1}}< \dot{E}_{\rm K} < 8.62 \times 10^{41} \ {\rm erg} \ {\rm s^{-1}}$,  
which is equivalent to $0.08\%-1.53\%$ of the bolometric luminosity, $L_{\rm bol}$. 
This result is consistent with other observed properties of ionized gas outflows, 
although it is lower than the theoretical predictions in AGN feedback models ($\sim5\%$), 
implying that ionized gas outflows traced by \OIII\ represent only a minor fraction of the total outflows ejected from the host galaxy. 
Given the asymmetric and radially distributed outflow signatures observed across the host galaxy 
within the limited field of view, the maximum distance the outflowing gas has traveled remains an open question. 
\end{abstract}

%\pagewiselinenumbers 

% ----------------------------------------------------- 
% 		INTRODUCTION
% ----------------------------------------------------- 
\section{Introduction}\label{intro}
Classical spectroscopic studies have been playing a critical role in investigating various properties of Active Galactic Nuclei (AGNs). 
Both fiber and slit spectroscopy have been used to acquire optical spectra of nearby AGNs, from which we can extract important information not only about supermassive black holes (SMBHs) and structure of AGNs but also about host galaxies (e.g., mass of SMBHs, Eddington accretion rate, metallicity, chemical enrichment, age, kinematics, and so on). 

Recent advances in integral field unit (IFU) spectroscopy have opened a new era that traditional approaches could not achieve. 
In particular, optical IFU observations from dedicated survey such as the SDSS-IV \citep{Blanton17} Mapping Nearby Galaxies at APO (MaNGA, \cite{Bundy15, Drory15, Law15, Yan16}) have identified `hidden' AGN signatures from spatially resolved regions of host galaxies \citep{Wylezalek18, Mezcua20, CanoDiaz22}.

AGN could be displaced from the center of its host galaxy due to a recent galaxy merger \citep{Comerford14, Barrows18, Bellovary19}. 
Another plausible explanation is that AGN may have recently been turned off \citep{Shapovalova10, McElroy16}, causing light echoes to travel large distances from the center and revealing their highly ionized relic signatures \citep{Keel12, Keel15, Schawinski15, BlandHawthorn19}.

By virtue of powerful capabilities of IFU observations, the ionized gas kinematics of sizable samples of AGNs has been intensely studied over the last decade. Studies have shown that kiloparsec-scale AGN-driven powerful outflows are prevalent \citep{f7_Liu13b, Harrison14, McElroy15, Rupke17, DecontoMachado22, Gatto24} not only in the centers of host galaxies but also in their extended regions, which were difficult to explore using traditional fiber and slit spectroscopy. These gas outflows are one of the key ingredients in the framework of the $\Lambda$ Cold Dark Matter ($\Lambda$CDM) cosmological model, as they regulate star formation, which is expected to align with the galaxy luminosity function.   
For this reason, IFU observations are essential not only for uncovering unknown and interesting features of AGNs but also as a crucial part of galaxy formation models.  

The Kyoto Okayama Optical Low-dispersion Spectrograph with optical fiber IFU (KOOLS-IFU, \cite{Matsubayashi19}) on Okayama 3.8m Seimei telescope \citep{Kurita20} has started its science operation since 2019. It has been actively used to explore unseen features from many fascinating Galactic \citep{Otsuka22, Otsuka23, Taguchi23, Namekata24} as well as extragalactic sources including AGNs \citep{Toba22, Hoshi24, Nagoshi24, Toba24}.

In this paper, we present the results of optical IFU observations of Mrk 766, one of the well-known nearby AGNs 
classified as a narrow-line Seyfert 1 galaxy (NLS1, \cite{Osterbrock85}), conducted for the first time using the KOOLS-IFU on the Seimei telescope. 
This paper is organized as follows. 
In Section \ref{DandA}, we describe the data we acquired and its analysis. 
In Section \ref{RandD}, we present the results along with measured quantities and discuss our findings, including important caveats. 
Finally, we briefly summarize our results in Section \ref{S}.   
We assume a cosmology with $h = 0.70$, $\Omega_{M} = 0.30$, and $\Omega_{\Lambda}=0.70$ throughout this work.

% ----------------------------------------------------- 
% 		TABLE 1: general properties
% ----------------------------------------------------- 
\begin{table}
  \tbl{Basic properties of Mrk766.\footnotemark[$*$] }{%
  \begin{tabular}{cccccc}
      \hline
      Name & RA 		& Dec 		& Redshift & V mag  & AGN type \\ 
      	        &  (J2000.0) 	& (J2000.0) 	&  		&  		&  \\       
       (1)     &  (2) 		& (3)			& (4)		& (5)		&  (6) \\   
      \hline
      Mrk 766 & 12:18:26.51 & +29:48:46.58 & 0.01292423 & 13.6 & NLS1 \\
      \hline
    \end{tabular}}\label{tab:tab1}
\begin{tabnote}
\footnotemark[$*$] Columns: 
(1) Object name. 
(2)-(3) Right ascension (RA) and declination (Dec) (J2000.0). 
(4) Spectroscopic redshift \citep{Koss_DR2_catalog}. 
(5) V-band magnitude \citep{Veron_Cetty10}. 
(6) AGN type \citep{Osterbrock85}.  \\ 
\end{tabnote}
\end{table}

% ----------------------------------------------------- 
% 		FIGURE: Mrk766 SDSS image
% ----------------------------------------------------- 
\begin{figure}
 \begin{center}
  \includegraphics[width=8cm]{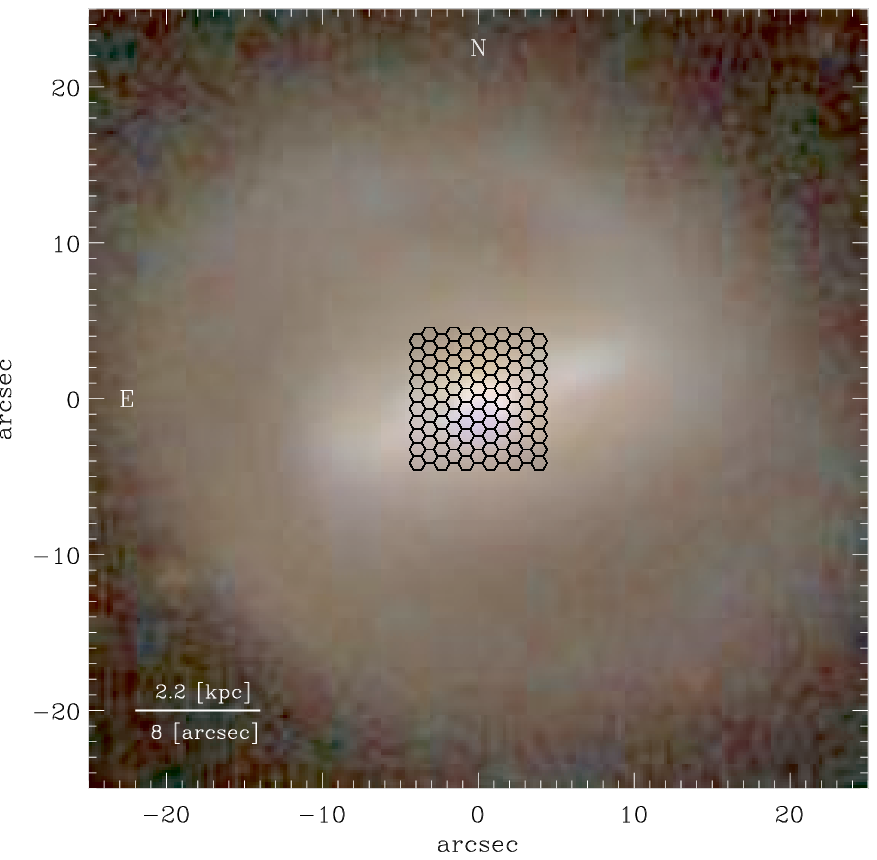}
 \end{center}
\caption{SDSS {\it gri} color composite image of Mrk 766, overlaid with the KOOLS-IFU arrays (black hexagons, 8.4$\arcsec$$\times$8$\arcsec$).  
}\label{fig:fig_sdss}
\end{figure}

% ----------------------------------------------------- 
% 		DATA AND ANALYSIS
% ----------------------------------------------------- 
\section{Data and analysis}\label{DandA}

\subsection{Target and observations}\label{targetobservations}

We chose Mrk 766 (also known as NGC 4253, table \ref{tab:tab1}) for this work because it has been widely studied across multiwavelength regime, 
from X-rays \citep{Miller07, Risaliti11, Tombesi12, Yamada24} to near-infrared (NIR, \cite{RodriguezArdila05, Schonell14, Riffel23}) and radio \citep{Kukula95}. 
\citet{GonzalezDelgado96} reported the prominent emission lines in the optical and NIR bands using long-slit spectroscopy, 
indicating the presence of outflows originating from the nucleus. 

Furthermore, as a part of the optical follow-up spectroscopic survey of the all-sky hard X-ray selected AGNs \citep{Oh18}, 
BAT AGN Spectroscopic Survey (BASS\footnote{https://bass-survey.com}, \cite{Koss17}) presented  
optical emission-line properties of Mrk 766 \citep{Oh22} including the mass of the SMBH (${\rm log}M_{\rm BH}=6.82_{-0.10}^{+0.07}$ $M_{\odot}$, \cite{MejiaRestrepo22}). 
Earlier, \citet{Woo02} and \citet{Bentz09} measured ${\rm log}M_{\rm BH}$ to be $6.54$ $M_{\odot}$ and $6.25_{-0.69}^{+0.28}$ $M_{\odot}$, respectively. 
Given that the errors in measuring black hole mass are dominated by the intrinsic spread of virial black hole mass estimates, 
which is on the order of 0.5 dex, the reported measurements are in agreement with each other.
The measurements of ${\rm log}M_{\rm BH}$ and ${\rm log}L_{\rm bol}=43.75$ \ergs\ by \citet{Koss_DR2_catalog} yield $-1.25$ for ${\rm log}(L/L_{\rm Edd})$. The bolometric luminosity by \citet{Koss_DR2_catalog} was calculated using the X-ray ($14-150$ keV) intrinsic luminosity (\cite{Ricci17}, $\kappa=8$), which is equivalent to a $2-10$ keV  with a conversion factor of 20 \citep{Vasudevan09} assuming $\Gamma=1.8$.

The observational data were taken using the KOOLS-IFU instrument operating on the Seimei Telescope under the programme 24A-K-0013 (PI: Y. Ueda). 
The KOOLS-IFU consists of 110 fibers\footnote{http://www.o.kwasan.kyoto-u.ac.jp/inst/p-kools/inst-info/} 
with a total field of view (FoV) of $8.4$$\arcsec$ by $8$$\arcsec$, 
which corresponds to approximately $2$ kpc $\times$ $2$ kpc (figure~\ref{fig:fig_sdss}).
We used the VPH 495 and VPH 683 grisms which have spectral resolution of ${\rm R(=\lambda/\Delta\lambda})=1500-2000$. The observation log is summarized in table~\ref{tab:tab2}.

% ----------------------------------------------------- 
% 		FIGURE: fitting example
% ----------------------------------------------------- 
\begin{figure*}
 \begin{center}
  \includegraphics[width=0.95\textwidth]{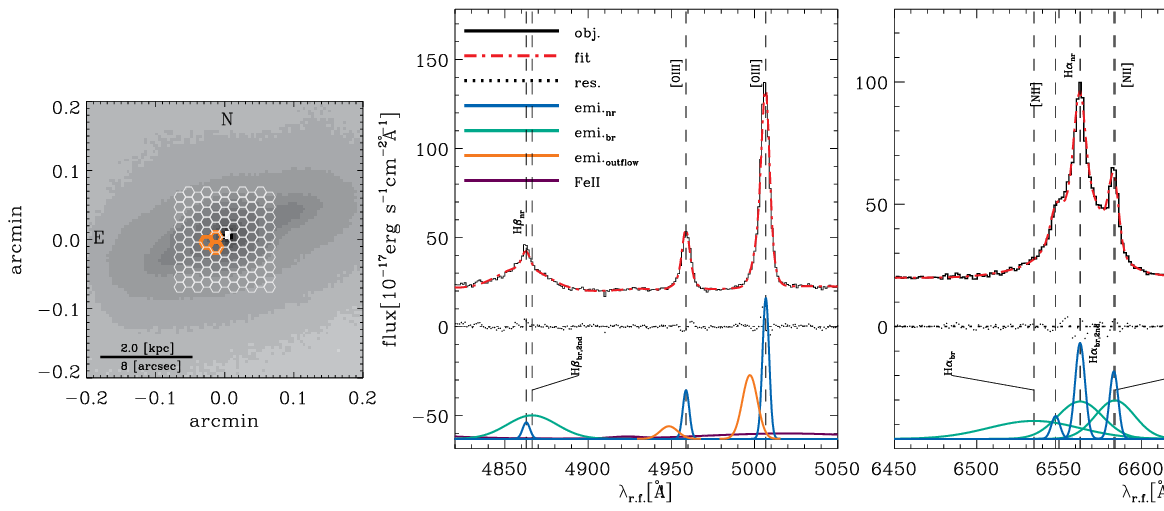}
 \end{center}
\caption{Example spectral line fitting results with the KOOLS-IFU arrays. 
The left panel shows the r-band image of Mrk 766, obtained from Pan-STARRS1 \citep{Chambers16}, overlaid with the KOOLS-IFU arrays. 
The region from which the spectrum is extracted is marked with orange hexagons. 
The middle and right panels present the detailed spectral fits. 
The black line represents the observed spectrum in the rest frame, while the red dashed-dotted line indicates the best fit. 
The blue Gaussians represent narrow emission-line components, 
and the green Gaussians represent broad emission-line components. 
Gas outflow components and FeII templates are depicted with orange and purple lines, respectively. 
The decomposed models are shown with arbitrary offsets for clarity. 
The dots in both panels represent the residuals. 
}\label{fig:fit_example}
\end{figure*}

% ----------------------------------------------------- 
% 		TABLE 2: Obs. log
% ----------------------------------------------------- 
\begin{table}
  \tbl{Observation log.}{%
  \begin{tabular}{ccccc}
      \hline
       		&  		& Exp.time & Exp.time 	&  \\ 
      Name	&  Date 	& VPH495 &  VPH683	& Stardard star \\       
      \hline
      Mrk 766 & 2024.04.13 & 500 $\times$ 3\footnotemark[$*$] & 500 $\times$ 3\footnotemark[$*$] & HR5191 \\
      \hline
    \end{tabular}}\label{tab:tab2}
\begin{tabnote}
\footnotemark[$*$]The total integration time is 1500 sec for each grism.
\end{tabnote}
\end{table}

% ----------------------------------------------------- 
% 		DATA REDUCTION
% ----------------------------------------------------- 
\subsection{Data reduction}\label{datareduction}
We reduced the data using the standard reduction pipeline\footnote{http://www.o.kwasan.kyoto-u.ac.jp/inst/p-kools/reduction-201806/index.html} for KOOLS-IFU, 
which employs the Image Reduction and Analysis Facility (IRAF, \cite{Tody86, Tody93}). 
For spectrum extraction, flat-fielding, and wavelength calibration, we used the Hydra package \citep{Barden94, Barden95}. 
We calibrated the wavelength using arc lamp frames (Ne and Hg). 
Absolute flux calibration of the spectrum was performed using standard star frames. 
To account for background sky levels, we took sky frames separately and subtracted them from the data frames. 
Considering the typical seeing at the site ($1.2\arcsec-1.4\arcsec$) and the field of view of each fiber, 
which has a regular hexagonal shape with a radius of 0.42$\arcsec$, 
we combined data obtained from three adjacent fibers and present the results throughout this work 
(see left panel in Fig~\ref{fig:fit_example}).  
As a result, we present data extracted from 98 out of a total of 110 fibers, of which 12 fibers located on the outskirts of the full arrays are not used.

% ----------------------------------------------------- 
% 		SPECTRAL FITTING 
% ----------------------------------------------------- 
\subsection{Spectral fitting}\label{spectralfitting}

We performed the spectral line fitting of the KOOLS-IFU data following the detailed procedures outlined by \citet{Sarzi06} and \citet{Oh11}, 
which have been extensively used for various types of spectra \citep{Oh15, Oh17, Rey21, Oh19, Oh22}. 
We first deredshifted the extracted spectra and corrected them for Galactic foreground extinction (${\rm E(B-V)}=0.0197$, \cite{Schlegel98}) 
using the dust attenuation curve of \citet{Calzetti00}. 
We fitted the stellar continuum using the penalized pixel fitting method (\texttt{pPXF}, \cite{Cappellari04}) 
which employs the stellar population models \citep{Bruzual03} and the MILES empirical stellar libraries \citep{SanchezBlazquez06}. 
We masked regions where emission lines are expected to be present using masks with widths of 1200 \kms. 
When broad lines are observed in the Balmer series, wider masks are employed to accommodate at least several thousand \kms\ of broadening. 
In addition to nebular emission lines that could potentially affect continuum fitting, 
we masked skylines (5577\AA, 6300\AA, and 6363\AA) as well as NaD $\lambda\lambda$5890, 5896 absorption lines during the process using the same line widths. 

After fitting the stellar continuum, we used the \texttt{gandalf} code \citep{Sarzi06} to simultaneously match the stellar continuum and emission lines. 
Emission lines are modeled to be Gaussian using either single or multiple templates (e.g., Balmer lines, \OIIIab, and \NIIab). 
We adopted the relative strengths of some emission lines based on atomic physics and the gas temperature (doublets, triplets, and Balmer lines, see table 1 in \cite{Oh11}). 

To resolve the shifts and widths of the Gaussian templates, 
we used the standard Levenberg-Marquardt optimization (\texttt{MPFIT} IDL routine; \cite{Markwardt09}). 
The stellar line-of-sight velocity dispersions derived from the earlier step were used to broaden the stellar templates in the fit. 
In the presence of broad Balmer line features, 
we applied additional Gaussian components with a full width at half maximum (FWHM) greater than 1000 \kms. 
For complex broad emission features presented in the \Hb\ and/or \Ha\ spectral regions, 
as demonstrated in the middle and right panels in figure~\ref{fig:fit_example}, 
we allowed multiple Gaussian components with shifted line centers, if necessary. 
We measured the error of the emission-line fluxes by resampling each emission line based on the 100 realizations with randomly added noise and measuring its 1$\sigma$ dispersion.

% ----------------------------------------------------- 
% 		RESULTS AND DISCUSSION
% ----------------------------------------------------- 
\section{Results and discussion}\label{RandD}

% ----------------------------------------------------- 
% 		FIGURE: luminosity maps
% ----------------------------------------------------- 
\begin{figure}
 \begin{center}
  \includegraphics[width=8cm]{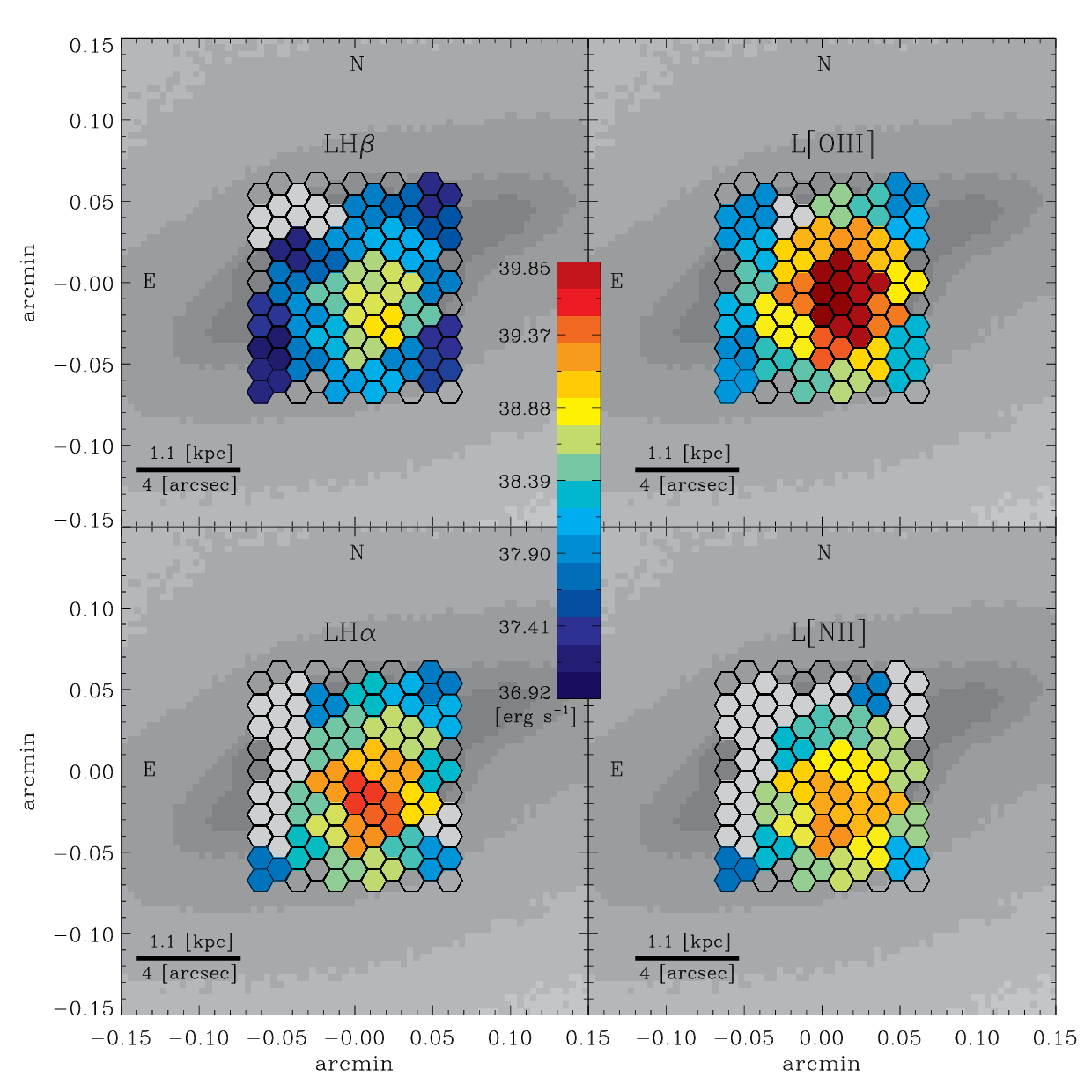}
 \end{center}
\caption{Emission-line luminosities (\Hb, \OIII, \NII, and \Ha, from top-left to bottom-left in clockwise). Gray hexagons denote the KOOLS-IFU fibers for which line luminosity could not be calculated due to poor SN. Emission-line luminosities are color-coded in log scale. 
}\label{fig:lum_maps}
\end{figure}

% ----------------------------------------------------- 
% 		FIGURE: BPT maps
% ----------------------------------------------------- 
\begin{figure}
 \begin{center}
  \includegraphics[width=0.48\textwidth]{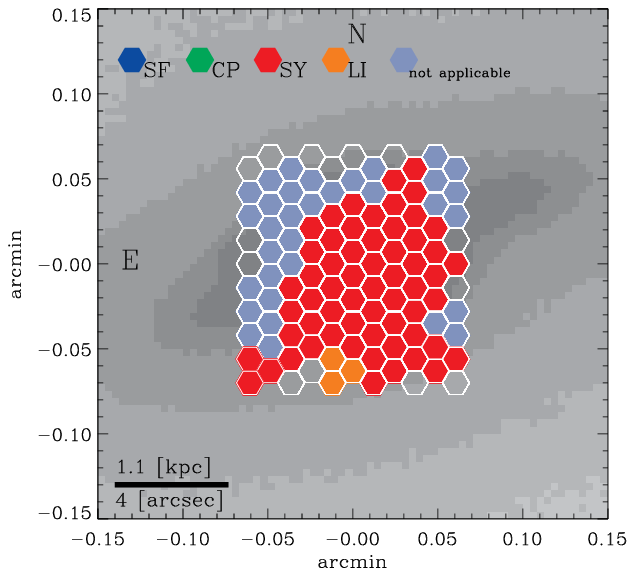}
 \end{center}
\caption{Spatially resolved map of BPT diagnostics diagram (\OIIIHb). 
Blue, green, red, and orange colors are designed to denote SF, composite, Seyfert, and LINER, respectively. 
Light blue hexagons indicate the fibers that could not be diagnosed either due to poor SN ratio ($<3$) or non-detection of the given emission lines. 
}\label{fig:BPT_maps}
\end{figure}

% ----------------------------------------------------- 
% 		SPATIALLY RESOLVED MAPS OF DERIVED PROPERTIES
% ----------------------------------------------------- 
\subsection{Spatially resolved maps of derived properties}\label{ssec:spatial}
We present spatially resolved maps of derived properties within the innermost $2$ kpc $\times$ $2$ kpc scale covered by the KOOLS-IFU arrays. 
Figure~\ref{fig:lum_maps} shows maps of the observed narrow emission-line luminosities (\Hb, \OIII, \Ha, and \NII). 
We did not correct for internal extinction for the sake of simplicity, as some fibers located in the outskirts exhibited poor SN ratios. 
This can be justified by the fact that the BPT diagram is, by definition, insensitive to dust extinction due to the small wavelength separation of the emission lines used.

In figure~\ref{fig:BPT_maps}, we present a spatially resolved map of the Baldwin, Phillips, and Terlevich (BPT) diagram \citep{Baldwin81}. 
We used both theoretical and empirical demarcation lines \citep{Kewley01, Kauffmann03, Kewley06, Schawinski07} to classify regions as star-forming region (SF), 
composite, Seyfert, and low-ionization nuclear emission-line region (LINER). 
Although we employed the full classification scheme outlined above, 
it is noteworthy that the majority of the regions observed by the KOOLS-IFU arrays 
within an approximately $2$ kpc $\times$ $2$ kpc scale are classified as Seyfert (red filled hexagons in figure~\ref{fig:BPT_maps})
according to the diagnostic diagram (\OIIIHb\ vs. \NIIHa), with some hexagons indicating LINERs in the outskirts.

% ----------------------------------------------------- 
% 		FIGURE: W80 & Vmed maps
% ----------------------------------------------------- 
\begin{figure*}
 \begin{center}
  \includegraphics[width=0.95\textwidth]{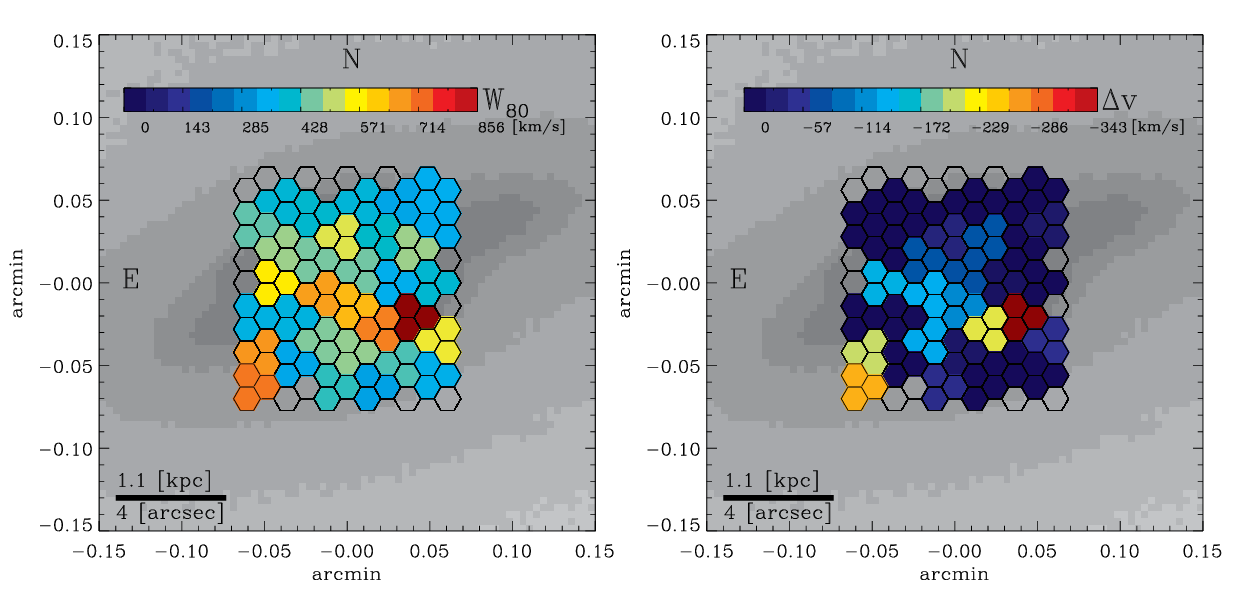}
 \end{center}
\caption{Spatially resolved map of $W_{\rm 80}$ (left) and $\Delta v$ (right). 
Asymmetric emission-line profiles with blue-shifted components are indicated by negative signs and color codes in the right panel. 
}\label{fig:W80_DeltaV_maps}
\end{figure*}

% ----------------------------------------------------- 
% 		NON-PARAMETRIC EMISSION LINE MEASURES
% ----------------------------------------------------- 
\subsection{Non-parametric emission line measures}\label{ssec:nonparametric}
In order to analyze strong gas outflow emission lines detected near \OIII\ (middle panel in figure~\ref{fig:fit_example}), 
we exploited non-parametric emission line measures following \citet{Harrison14} and \citet{McElroy16}. 
The non-parametric line width, $W_{\rm 80}$, which is the line width containing the central 80 per cent of the flux, is defined as follows, 
\begin{equation}
W_{\rm 80}=v_{\rm 90}-v_{\rm 10},
\end{equation}
where $v_{\rm 90}$ and $v_{\rm 10}$ are defined as the velocities at 90 and 10 per cent of the line flux, respectively. 

An asymmetric parameter, $\Delta v$, is defined as follows, 
\begin{equation}
\Delta v=\frac{v_{\rm 05}+v_{\rm 95}}{2}-v_{\rm med},
\end{equation}
where $v_{\rm 05}$ and $v_{\rm 95}$ are the velocities at 5 and 95 per cent of the line flux, respectively. 
$v_{\rm med}$ is the median velocity that bisects the total flux of the emission line, 
allowing us to trace the direction of the propagating outflow gas across the host galaxy. 

Figure~\ref{fig:W80_DeltaV_maps} clearly demonstrates that the innermost $2$ kpc $\times$ $2$ kpc region of Mrk 766 exhibits 
highly disturbed ionized gas kinematics, characterized by high-velocity (${>}500$ \kms, \cite{Wylezalek20}) 
blue-shifted component observed in the \OIII\ emission-line. 
An asymmetric emission-line profile, featuring a blue wing that reaches high velocities, is a definitive signature of an AGN-driven outflow \citep{Harrison16, Kakkad20}. 
Such extremely broad blue-shifted emission lines are highly unlikely to arise from normal rotational motions in galaxy kinematics \citep{VegaBeltran01} 
or supernovae-driven outflows \citep{Thacker06}. 
It should be noted that a similar, albeit larger, threshold in $W_{\rm 80}$ (${>}600$ \kms) was used in previous works studying quasars (e.g., \cite{Kakkad20}). 

The right panel in figure~\ref{fig:W80_DeltaV_maps} provides insight into the distribution of outflowing gas components 
across the host galaxy within the observed region. 
It is evident that the radially distributed AGN-driven outflow gas is shaping the map of $\Delta v$ in figure~\ref{fig:W80_DeltaV_maps}, 
extending in the south-east and south-west directions from the center, 
and is marginally misaligned with the galaxy bar. 

\begin{figure*}
 \begin{center}
  \includegraphics[width=0.95\textwidth]{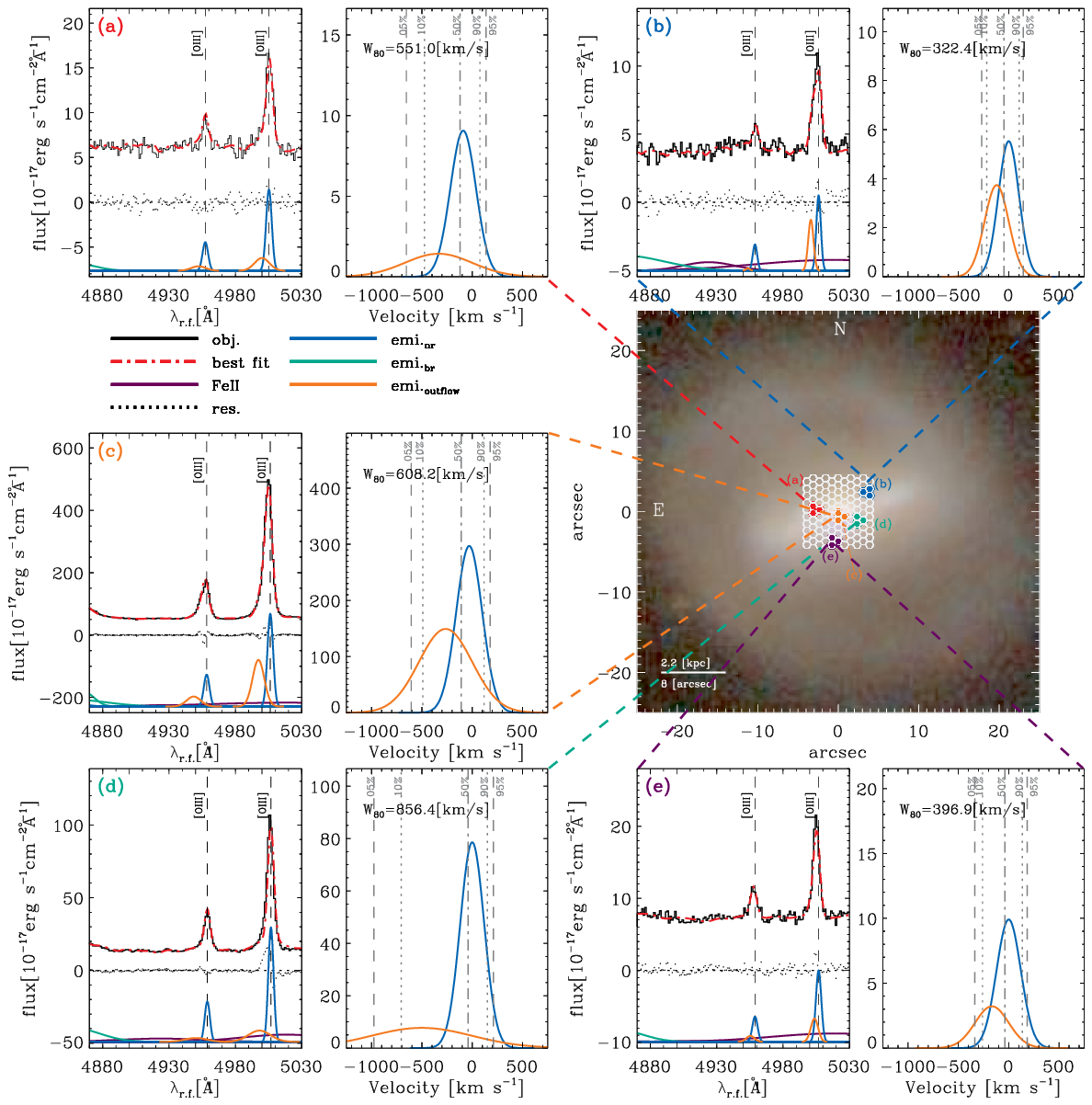}
 \end{center}
\caption{AGN-driven outflow gas components detected within the KOOLS-IFU arrays across the host galaxy. 
The vertical dark grey lines in the panels indicating velocity denote $v_{\rm 05}$, $v_{\rm 10}$, $v_{\rm 50}$, $v_{\rm 90}$, and $v_{\rm 95}$ (see Section~\ref{ssec:nonparametric}). 
Color-filled hexagons and the corresponding dashed lines, overlaid on the SDSS \textit{gri} composite image, illustrate the locations from which the spectrum is extracted. 
In the case of low Gaussian amplitude over noise ratio ($<3$) in emission lines, red labels are used. 
The format of the surrounding panels is the same as that of figure~\ref{fig:fit_example}.
}\label{fig:image_with_spec_5panels}
\end{figure*}

This trend is prominent in figure~\ref{fig:image_with_spec_5panels}. 
Notably, the AGN-driven outflow gas is not strongest at the center of galaxy, where the AGN of Mrk 766 is located. 
The outflow gas velocity at the center exceeds 500 \kms\ (orange filled hexagons and panel (c) in figure~\ref{fig:image_with_spec_5panels}). 
The AGN-driven outflows are weaker in the north-east (panel (a)) and north-west (panel (b)) directions compared to the central region. 
In contrast, the strongest outflow component ($W_{\rm 80}\approx860$ \kms) is identified in the south-west direction (panel (d)), 
along with a strong outflow component found at the edge of the KOOLS-IFU arrays in the south-east direction.

These AGN-driven outflows near \OIII\ in a kpc-scale represent new findings in optical spectroscopy. 
Note that earlier work by \citet{Oh22} demonstrated the AGN nature of Mrk 766 through emission-line diagnostics and the presence of broad Balmer features 
via optical long-slit spectroscopy, which lacks sufficient spectral resolution and the capability to detect outflow signatures from regions beyond the center of the host galaxy.

The detection of AGN-driven outflows on different physical scales has been reported across multiwavelength regimes, 
including X-rays, optical, NIR, and radio. 
From the analysis of X-ray data, ultrafast outflows on sub-parsec scales have been discussed \citep{Tombesi12, Yamada24}. 
\citet{Yamada24}, in particular, presented the ultrafast outflow velocities exceeding $10,000$ \kms, detected over a 10-year period from 2005 to 2015. 
\citet{Fischer13} noted the presence of highly blue-shifted velocity components in Mrk 766 using Hubble Space Telescope (HST) Space Telescope Imaging Spectrograph (STIS), 
concluding that the outflow cannot be resolved into individual components.
In the NIR, \citet{Schonell14} and \citet{Riffel23} showed evidence of an outflow using the Near Infrared Integral Field Spectrograph (NIFS) at the Gemini Telescope. 
\citet{Schonell14} concluded that the outflow is present at a position angle of approximately $135 \degree$, 
which is consistent with the results of the present study, 
where the [FeII] velocity dispersion is increased and coincides with the location of the radio jet \citep{Kukula95}. 
It is worth noting that the NIR studies using NIFS covered a smaller region of approximately $900$ pc $\times$ $900$ pc, 
which corresponds to about 20\%\ of the area observed in this study. 

% ----------------------------------------------------- 
% 		FIGURE: highlight 
% ----------------------------------------------------- 

% ----------------------------------------------------- 
% 		OUTFLOW PROPERTIES: MASS, RATE, KINETIC POWER
% ----------------------------------------------------- 
\subsection{Outflow properties}\label{ssec:outflowproperties}
The mass of the ionized outflowing gas can be expressed as follows, 
\begin{equation}
M_{\rm out} = 5.3 \times 10^{7} \frac{ L_{44}{\rm ([OIII])} } {n_{\rm e} 10^{\rm \left[O/H\right]} } M_{\odot},
\end{equation}
where $ L_{44}{\rm ([OIII])} $ is the extinction-corrected luminosity of the outflow component detected in \OIIIb, measured in units of $10^{44}$ \ergs, 
$n_{\rm e}$ is the electron density in the outflowing gas, expressed in units of $10^{3}$ ${\rm cm^{-3}}$, 
and $10^{\rm \left[O/H\right]}$ represents the oxygen abundance in Solar units (see Appendix B in \citet{CanoDiaz12} for more details). 
Assuming the electron density to be $50$ ${\rm cm^{-3}}$ $<n_{\rm e}<$ $1000$ ${\rm cm^{-3}}$, we find the mass of the ionized outflowing gas to be $10^{4.65}-10^{5.95}M_{\odot}$. 
Under the same assumptions regarding the oxygen abundance and the same range for $n_{\rm e}$, 
we find the mass of the ionized outflowing gas to be $10^{4.52}-10^{5.82}M_{\odot}$ using the formulae from \citet{Carniani15}. 
The electron density derived from the \SIIab\ is known to be significantly lower than that measured by several other recent methods, such as those using auroral lines and the ionization parameter. 
Indeed, determining the electron density is a highly uncertain and complex issue, as it varies as a function of luminosity, ionization parameter, and distance from the AGN \citep{Davies20}. 
Therefore, we leave the values of the ionized outflowing gas as a range that varies with the choice of $n_{\rm e}$, rather than assuming a constant. 

The mass of the ionized outflowing gas presented in this study is most likely a lower limit for two reasons. 
First, the limited field of view of the KOOLS-IFU instrument corresponds to approximately 2 kpc $\times$ 2 kpc, covering the central part of the host galaxy. 
We do not have additional data outside of the given FoV, which could contribute to the total ionized outflowing gas. 
Second, and more importantly, the ionized outflow gas traced by \OIII\ is a subset of the total outflow.

The outflow rate of ionized gas can be denoted as follows \citep{CanoDiaz12}, 
\begin{equation}
\dot{M}_{\rm out} = 164 \frac{ L_{44}{\rm ([OIII])}v} { n_{\rm e} 10^{\rm \left[O/H\right]}  R_{\rm kpc} } \ M_{\odot} \ {\rm yr}^{-1},
\end{equation}
where $v$ is the outflow velocity in units of 1000 \kms, and $R_{\rm kpc}$ is the radius of the outflowing region, in units of kpc. 
By adopting $v=1000$ \kms, $50$ ${\rm cm^{-3}}$ $<n_{\rm e}<$ $1000$ ${\rm cm^{-3}}$, and $R_{\rm kpc}=1$ kpc for simplicity, 
we find the ionized gas outflow mass rate to be $0.14-2.73$ $M_{\odot}{\rm yr}^{-1}$. 

\citet{Tombesi12} and \citet{Riffel23} reported $\sim$ $0.001-1$ $M_{\odot} {\rm yr}^{-1}$ and 
$\sim$ $0.3$ $M_{\odot} {\rm yr}^{-1}$ for $\dot{M}^{\rm out}$, respectively. 
Given the different bands (X-ray, NIR vs. optical), spectroscopic lines (Fe K, Br$\gamma$ vs. \OIII), and assumptions (fixed $n_{\rm e}$ vs. variable) used to measure $\dot{M}^{\rm out}$, 
it is difficult to directly compare them to each other. 
However, we would like to note that the inferred $\dot{M}^{\rm out}$ values are consistent within an order of magnitude. 

\begin{figure}
 \begin{center}
  \includegraphics[width=0.48\textwidth]{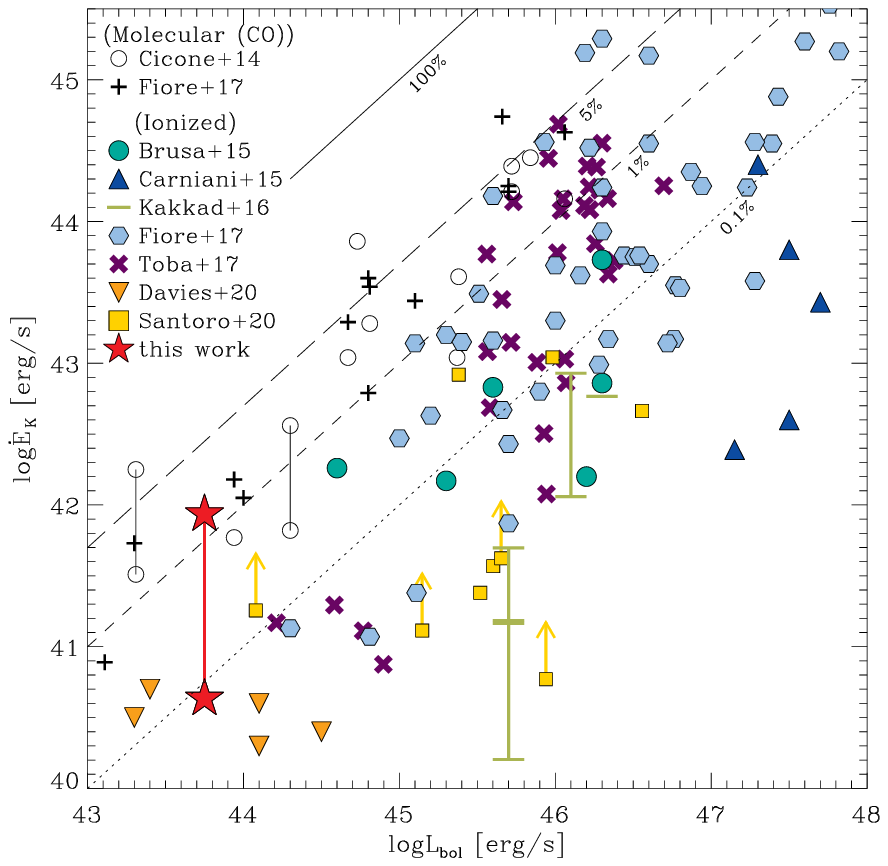}
 \end{center}
\caption{Kinetic power of ionized gas outflow, $\dot{E}_{\rm K}$ vs. bolometric luminosity, $L_{\rm bol}$.
The solid, long-dashed, dashed, and dotted lines represent the outflow kinetic power as  
$100\%$, $5\%$, $1\%$, and $0.1\%$ of the AGN bolometric luminosity, respectively.
Molecular outflows are presented by empty black circles and plus symbols \citep{f7_Cicone14, f7_Fiore17}\footnotemark[$*$], 
while the remaining colored data points correspond to ionized gas outflows \citep{f7_Brusa15b, Carniani15, f7_Kakkad16, f7_Fiore17, Toba17, Davies20, Santoro20}\footnotemark[$*$].
The measurements of Mrk 766 in this work are presented as red-filled stars, 
with a range constrained by $n_{\rm e}$.
}\label{fig:kinetic_power_vs_Lbol}
\footnotemark[$*$] \citet{f7_Fiore17} compiled the measurements from \citet{f7_Downes98, f7_Tacconi99, f7_Lonsdale03, f7_Davies04, f7_Dasyra06, f7_Nesvadba06, f7_Davies07, f7_Nesvadba08, f7_Reyes08, f7_Veilleux09, f7_Engel10, f7_Howell10, f7_Harrison12, f7_Maiolino12, f7_Feruglio13, f7_Liu13a, f7_Liu13b, f7_Cicone14,  f7_Genzel14, Harrison14, f7_Sun14, f7_Brusa15a, f7_Brusa15b, f7_Cicone15, f7_Cresci15, f7_Feruglio15, f7_Perna15a, f7_Perna15b, f7_Brusa16, f7_Kakkad16, f7_Wylezalek16, f7_Bischetti17, f7_Duras17}. 
\end{figure}

The kinetic power of the ionized gas components of the outflows can be expressed as,  
\begin{equation}
\dot{E}_{\rm K} = 5.17 \times 10^{43}\frac{ L_{44}{\rm ([OIII])}v^{3}} {n_{\rm e} 10^{\rm \left[O/H\right]} R_{\rm kpc}} \ {\rm erg} \ {\rm s}^{-1},
\end{equation}
following the same assumptions and notations as adopted above. 
We find the kinetic power ($\dot{E}_{\rm K}$) associated with the outflows to be 
$4.31 \times 10^{40} \ {\rm erg} \ {\rm s^{-1}}< \dot{E}_{\rm K} < 8.62 \times 10^{41} \ {\rm erg} \ {\rm s^{-1}}$. 
The measured $\dot{E}_{\rm K}$ corresponds to $0.08\%-1.53\%$ of $L_{\rm bol}$, under the given range of $n_{\rm e}$, 
in agreement with earlier theoretical works that predict the relationship between the kinetic energy released by the AGN 
and the energy required to induce the outflow ($\sim 0.1-5\%$, \cite{King05}). 
This is also consistent with the observational work of \citet{Tombesi12}, 
who showed that the typical ratios between the kinetic power of the outflows and the bolometric luminosities of 28 local AGNs 
are $\dot{E}_{\rm K}/L_{\rm bol}>0.3 \%$ for ultrafast outflows and 
$\sim 0.02 \%<\dot{E}_{\rm K}/L_{\rm bol}<0.8 \%$ for non-ultrafast outflows.
In figure~\ref{fig:kinetic_power_vs_Lbol}, 
we present the kinetic power of the ionized gas (colored symbols, \cite{f7_Brusa15b, Carniani15, f7_Kakkad16, f7_Fiore17, Toba17, Davies20, Santoro20})
and molecular gas (empty black circles and plus symbols, \cite{f7_Cicone14, f7_Fiore17}), 
along with the measurement for Mrk 766 conducted in this work (red-filled stars). 
The data for Mrk 766 are well-aligned in terms of the coupling kinetic efficiencies $(\dot{E}_{\rm K}/L_{\rm bol})$ 
with other observed ionized gas outflows. 
This is somewhat lower than the theoretical predictions \citep{King10, Zubovas12}, which are about $\sim5\%$ of $(\dot{E}_{\rm K}/L_{\rm bol})$.
However, such a discrepancy can be interpreted as a natural consequence that  
our measurements of $(\dot{E}_{\rm K}/L_{\rm bol})$ likely trace only a small fraction of the total outflowing gas from Mrk 766. 

Lastly, we compare the mass outflow rate in the form of ionized gas outflows 
with the accretion rate feeding the central AGN. 
The mass accretion rate can be expressed as follows, 
\begin{equation}
\dot{m} = \frac{L_{\rm bol}}{\eta c^{2}} \ M_{\odot} \ {\rm yr}^{-1},
\end{equation}
where $L_{\rm bol}$ is the bolometric luminosity, 
$\eta$ is the radiative efficiency,   
and $c$ is the speed of light. 
We obtain a mass accretion rate of $\dot{m}\sim0.01$ $M_{\odot} \ {\rm yr}^{-1}$, assuming $\eta=0.1$, 
which is a typical value 
(e.g., \cite{Soltan82, Chokshi92, Yu02, Marconi04, Cao08, Ueda14})
in a geometrically thin and optically thick standard accretion disc model.
The mass outflow rate $(\dot{M}_{\rm out})$ we obtained is about 10 times larger $(n_{\rm e}=1000$ ${\rm cm^{-3}})$, 
or about 300 times larger $(n_{\rm e}=50$ ${\rm cm^{-3}})$, than the mass accretion rate $(\dot{m})$.  
Assuming that the mass outflow rate is constant across different physical scales, from sub-parsec to kilo-parsec, 
this implies that most of the mass is ejected in the form of ionized gas outflows,  
while only a minor fraction of the gas ($0.3\%-10\%$) is involved in the accretion onto the SMBH (e.g., \cite{Izumi23}).

% ----------------------------------------------------- 
% 		FIGURE: Kinetic Power of ionized gas outflow vs. Bolometric luminosity
% ----------------------------------------------------- 

\subsection{Caveats}\label{ssec:caveats}
As discussed in Section~\ref{ssec:outflowproperties}, 
the mass, rate, and kinetic power of AGN-driven ionized gas outflows 
vary significantly depending on the choice of $n_{\rm e}$, 
which we did not attempt to constrain in this work. 
Given the current data obtained from the KOOLS-IFU observations 
and the complex nature of $n_{\rm e}$ as discussed in \citet{Davies20}, 
an investigation of realistic values for $n_{\rm e}$ is beyond the scope of this study. 

Furthermore, the measured quantities ($M_{\rm out}$, $\dot{M}_{\rm out}$, and $\dot{E}_{\rm K}$) 
should be regarded as lower limits for the total outflowing gas, 
since the detected ionized gas component likely presents only a minor fraction of the total outflows across the host galaxy.

Additionally, we emphasize that we confirmed the existence of AGN-driven ionized gas outflows 
up to approximately 1 kpc from the center of the galaxy. 
This does not necessarily restrict the regions to which the outflow gas has reached,  
as the presented observational results are limited to the area covered by the KOOLS-IFU. 
Instead, it should be interpreted as our observation of powerful outflow gas traveling toward the outer regions of the host galaxy, 
as captured by the KOOLS-IFU instrument, 
which has a limited field of view compared to the size of Mrk 766. 
Whether stronger outflows are moving at greater distances remains unknown, 
and further comprehensive observational studies are required to better constrain the properties of AGN-driven outflow gas.

\section{Summary}\label{S}
We have presented optical IFU observations of innermost regions of Mrk 766, covering approximately 2 kpc $\times$ 2 kpc, 
using the KOOLS-IFU on the Okayama 3.8m Seimei telescope. 
Through multi-component spectral line fitting, we decomposed narrow emission lines as well as broad lines, 
detecting powerful outflowing gas in \OIIIb\ with velocities exceeding 500 \kms. 

Non-parametric line width ($W_{\rm 80}$) and the asymmetry parameter ($\Delta v$) indicate that 
the central region of Mrk 766 is kinematically disturbed. 
We identified radially distributed strong AGN-driven outflows in the south-east and south-west, while the northern side from the center is weaker. 

We estimated the mass of ionized outflowing gas to be $10^{4.65-5.95} M_{\odot}$, 
assuming an electron density in the range of $50$ ${\rm cm^{-3}}$ $<n_{\rm e}<$ $1000$ ${\rm cm^{-3}}$. 
Under the same assumption, we estimated the outflow rate of ionized gas to be $0.14-2.73$ $M_{\odot} {\rm yr}^{-1}$.
We measured the kinetic power of the ionized outflow gas 
($4.31 \times 10^{40} \ {\rm erg} \ {\rm s^{-1}}< \dot{E}_{\rm K} < 8.62 \times 10^{41} \ {\rm erg} \ {\rm s^{-1}}$),
which corresponds to a coupling efficiency of $0.08\%-1.53\%$.
The discrepancy between the observations and the theoretical predictions is likely a natural consequence of the fact that 
the observations used in this work trace only a small fraction of the total outflowing gas. 
The mass outflow rate in Mrk 766 is much larger ($10-300$ times) than the mass accretion rate, 
suggesting that at most $10\%$ of the inflow is accreted by the SMBH, 
under the assumption that the mass outflow rate is constant across the different physical scales. 
Finally, we emphasize that the question of how far the outflowing gas has traveled remains uncertain.

\begin{ack}
K.O. acknowledges support from the Korea Astronomy and Space Science Institute under the R\&D program (Project No. 2025-1-831-01), supervised by the Korea AeroSpace Administration, and the National Research Foundation of Korea (NRF) grant funded by the Korea government (MSIT) (2020R1C1C1005462, RS-2025-00553982).
This work is also supported by the Grant-in-Aid for Scientific Research 20H01946 (Y.U.), 
23K13154 (S.Y.), 23K22537 (Y.T.), 20K14521 (K.I.), 23K13147 (A.T.), 21J13894 (S.O.), 24K17104 (S.O.), 22KJ1990 (R.U.), and 23H04894 (K.M.).
S.Y. is grateful for support from the RIKEN Special Postdoctoral Researcher Program. 
This work is partly supported by the Optical and Infrared Synergetic Telescopes for Education and Research (OISTER) program funded by the MEXT of Japan (K.I.).
A.T. is partly supported by the Kagoshima University postdoctoral research program (KU-DREAM). 
We thank the staff of Okayama Observatory, Kyoto University, and Okayama Branch Office, NAOJ, NINS, for their help in the KOOLS-IFU observations.
\end{ack}

\bibliography{references}{}
\bibliographystyle{pasj}

\end{document}